\documentclass[%
 reprint,
 amsmath,amssymb,
 aps,
]{revtex4-1}

\usepackage{graphicx}
\usepackage{epstopdf}
\usepackage{dcolumn}
\usepackage{bm}
\usepackage{subfigure}
\usepackage{upgreek}

\begin{document}

\preprint{APS/123-QED}

\title{Atomic inner-shell radiation seeded free-electron lasers}

\author{Han Zhang$^{1,2}$}
\author{Kai Li$^{2,3}$}
\author{Jiawei Yan$^{2,3}$}
\author{Haixiao Deng $^{2,}$}%
 \email{denghaixiao@sinap.ac.cn}
\author{Baoyuan Sun$^{1,}$}\email{sunby@lzu.edu.cn}
 \affiliation{%
 $^1$School of Nuclear Science and Technology, Lanzhou University, Lanzhou 730000, China\\
 $^2$Shanghai Institute of Applied Physics, Chinese Academy of Sciences, Shanghai 201800, China\\
 $^3$University of Chinese Academy of Sciences, Beijing 100049, China
}%

%



\date{\today}

\begin{abstract}
 In order to effectively improve the output quality of X-ray free electron laser (XFEL), we theoretically propose an XFEL scheme seeded by atomic inner-shell laser. As well known, an atomic inner-shell laser based on neutral atoms and pumped by an XFEL has been experimentally demonstrated, which produced sub-femtosecond X-ray pulses with increased temporal coherence. It shows that, by using the inner-shell laser as a seed to modulate the electron bunch, very stable and almost fully-coherent short-wavelength XFEL pulses can be generated. The proposed scheme holds promising prospects in X-ray wavelengths, and even shorter.
\begin{description}
\item[PACS numbers]
41.60.Cr
\end{description}
\end{abstract}

\pacs{Valid PACS appear here}
\maketitle

\section{\label{sec:level1}Introduction}

The generation of lasers has brought great benefits to physics, chemistry and biology researches. These researches have aimed to produce short-wavelength lasers that generate coherent X-rays \cite{matthews1985demonstration,suckewer1985amplification}. The shorter the wavelength, the better the imaging resolution of the laser and the shorter the pulse duration, leading to better temporal resolution in probe measurements. Recently, X-ray free electron lasers (XFEL) based on self-amplified spontaneous emission (SASE) \cite{kondratenko1980generating,Bonifacio1984Collective} have made it possible to generate a hard X-ray laser (that is, photon energy is of the order of ten kiloelectronvolts) in an {\aa}ngstr\"{o}m-wavelength regime \cite{P2010First,Ishikawa2012A}. Nevertheless, this process starts from the electron beam shot noise and creates light pulses of limited temporal coherence \cite{vartanyants2011coherence}. As well known, both narrow bandwidth and excellent longitudinal coherence are very important for ultrafast X-ray spectroscopy \cite{young2010femtosecond}, X-ray quantum optics \cite{glover2012x}, and fundamental physics applications. In order to obtain fully coherent X-ray light pulses, a variety of new lasing schemes have been proposed. For example, direct-seeding \cite{togashi2013extreme}, self-seeding \cite{geloni2011novel,amann2012demonstration,Ratner2015Experimental}, high-gain harmonic generation \cite{Yu1991Generation,yu2000high,yu2003first}, echo-enhanced harmonic generation \cite{stupakov2009using}, phase-merging enhanced harmonic generation \cite{deng2013using,feng2014phase}, X-ray free-electron laser oscillator \cite{kim2008proposal,dai2012proposal,Li2017Gain}, in particular the self-seeding method is the most widely used scheme in hard XFEL facilities. However, the large pulse-to-pulse energy jitter and limited spectral brightness enhancement prevent it from further developments.

An alternative approach to create a pulse source in the X-ray regime is to use a laser to pump an atomic inner-shell X-ray laser(XRL). The first inner-shell XRL in the kiloelectronvolt regime was proposed in 1967 and is based on establishing a population inversion by rapid photo-ionization of an inner-shell electron \cite{duguay1967some}. The photo-ionization scheme was extensively studied in theory for different gain materials and pump sources in the X-ray regime \cite{kapteyn1992photoionization,axelrod1976inner,eder1994tabletop,strobel1993innershell}. However, the lack of sufficiently fast and intense X-ray sources has so far precluded the realization of the photo-ionization scheme in the X-ray regime. The introduction of XFEL makes it possible to pump new atomic inner-shell XRL \cite{lan2004photopumping,zhao2008x,rohringer2009atomic,rohringer2009atomic,darvasi2014optical} with ultrashort pulse duration, extreme spectral brightness and full temporal coherence. Experimentally, in 2012, An atomic laser based on neon atoms and pumped by a soft XFEL has been achieved at a wavelength of 14.6 {\aa}ngstr\"{o}ms in Linac Coherent Light Source \cite{rohringer2012atomic}; in 2015, the SPring-8 Angstrom Compact Free Electron Laser use a copper target and report a hard X-ray atomic inner-shell laser operating at a wavelength of 1.5 {\aa}ngstr\"{o}ms \cite{yoneda2015atomic}. Experimental results show that temporal coherence was greatly improved in the pumped neon (Ne) and copper (Cu) medium using XFEL.

In this work, we propose a scheme that combines self-seeding method and atomic inner-shell XRL. First, we use SASE XFEL as a pump source to pump a given neutral atom, which can produce a longitudinal coherent inner-shell XRL. Then this inner-shell XRL is used as a seed laser to modulate the electron beam, and generate a fully-coherent, short-wavelength and stable XFEL. The principle of our scheme, including the seeded XFEL and X-ray lasing in core excited atom are described in section \textrm{II}. The small-signal gain and the propagation of atomic inner-shell XRL are discussed in section \textrm{III}. The results of the seeded XFEL is illustrated in section \textrm{IV}, and finally, conclusions are drawn in section \textrm{V}.

\begin{figure*}
  \centering
  \subfigure{\includegraphics[width=16cm]{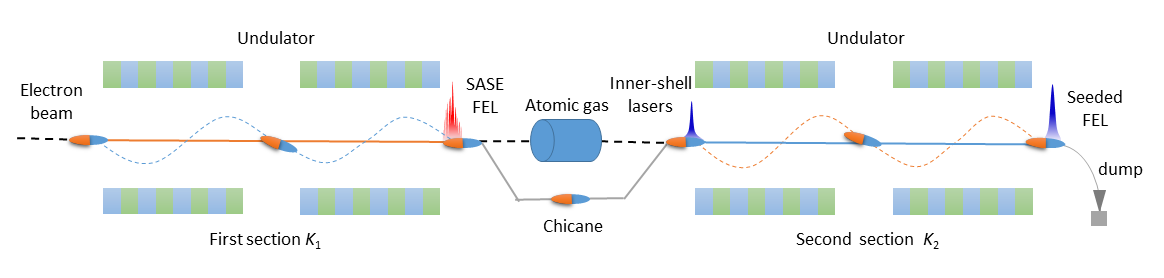}}
 \caption{\label{figure1} (color online).  \textbf{The seeding scheme:} The electron bunch travels off-axis in the dechirper experiencing a strong transverse head-tail kick, electron bunch slices and trajectories are represented in blue for the head and orange for the tail. While passing through the first part of the undulator, the electron bunch orbit is steered to have the bunch tail travelling straight. A saturated SASE XFEL pulse at energy $E_1$ is generated on the bunch tail in the first undulator section tuned to $K_1$. After that this XFEL pump a given neutral atomic gas and it can produce an extremely longitudinal coherent inner-shell XRL at energy $E_2$. Then this inner-shell XRL is used as seeding laser to modulate the electron bunch in the head, The bunch orbit is steered, the bunch head is travelling straightly in the second undulator section. The second undulator section tuned to $K_2$, its resonance energy is $E_2$. A fully saturated light pulse at energy $E_2$ is generated on the head.}
\end{figure*}

\section{The principles of scheme}

  A scheme that exploits the combination of an atomic inner-shell laser and a self-seeding method is presented in Fig.~\ref{figure1}. Here the fresh-slice technique was exploited \cite{lutman2016fresh}, where different temporal slices of an electron bunch lase to saturation in separate undulator sections. The electron bunch travels off-axis in the dechirper experiencing a strong transverse head-tail kick. While passing through the first part of the undulator, the electron bunch tail is travelling straight. A saturated SASE XFEL pulse at energy $E_1$ is generated. After that this XFEL is used to ionize a given neutral atom. It can produce an extremely longitudinal coherent inner-shell XRL at energy $E_2$ ($E_1>E_2$). Then this inner-shell XRL is used as seed laser to modulate the electron bunch in the head, at the same time, the bunch orbit is steered again and thus the bunch head travels straightly in the second undulator section. A fully saturated photon beam at energy $E_2$ is generated on the head in the second undulator section.

About the generation of atomic inner-shell XRL, it is based on atomic population inversion and driven by rapid \emph{K}-shell photo-ionization using pump sourse. This scheme was proposed as early as 50 years ago \cite{duguay1967some}, but has not been experimentally implemented due to the lack of a suitable X-ray pump source. The introduction of XFEL makes it possible to pump new atomic inner-shell XRL. XFEL is used to ionize the core electrons of neutral atoms to produce atomic population inversion, in which excited atoms are attenuated by Auger or radiation. Inner-shell photoionization happens on an ultrafast time scale of a few femtoseconds, the ion temperature in the plasma column is expected to remain close to room temperature during the time of amplification.

\begin{figure}
  \centering
  \includegraphics[width=8cm]{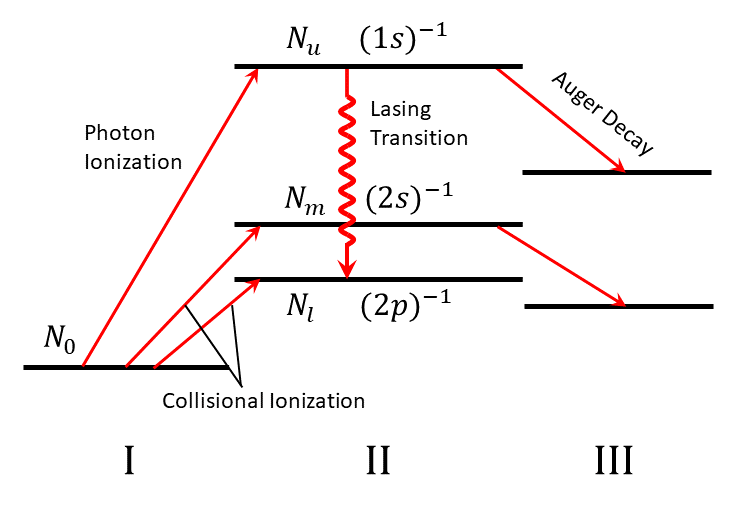}\\
  \caption{(color online). Schematic diagram of the main inner-shell vacancy levels and atomic processes included in the pumping of inner-shell XRL.}\label{figure2}
\end{figure}

Fig.~\ref{figure2} shows the considered scheme with the main energy levels related to the inner-shell vacancies and the atomic processes of the direct inner-shell photoionization, Auger decay and the secondary electron collisional ionization included in the photo-ionization pumped atomic inner-shell XRL. The ground state of the neutral atom and the lasing-related lower and upper states are represented by 0, $l$, $u$, respectively. The inner-shell vacancies (1$s$)$^{-1}$ are preferentially produced by the X-ray photons with an energy just above the inner-shell ionization threshold. Such light source is now available on the XFEL. The X-ray photons maximize the photoionization cross section of the inner-shell electrons and minimize the secondary electrons production through the photoionization of outer-shell electrons. A population inversion results from the initial photoionization.

\section{Atomic inner-shell laser}

Nowadays, more and more XFEL facilities are under construction and operation all over the world, which can generate extremely ultra-short pulse and high intensity radiation. And currently, the first hard XFEL in China is under construction in Shanghai, namely Shanghai Coherent Light Facility (SCLF). SCLF plans to utilize the superconducting linear accelerator to produce quasi-CW relativistic electron beams. The main parameters of the electron beams are as follows: 8 GeV beam energy, 1.5 kA peak current, 0.4 mm-mrad normalized emittance, 0.01\% relative energy spread, and 20-200 pC bunch charge. As a high repetition rate machine, three FEL lines, i.e., FEL-\textrm{I}, FEL-\textrm{II} and FEL-\textrm{III} will be equipped simultaneously to cover the full photon range of 0.3-25 keV in the first phase of SCLF.

\begin{figure}
  \centering
  \includegraphics[width=8cm]{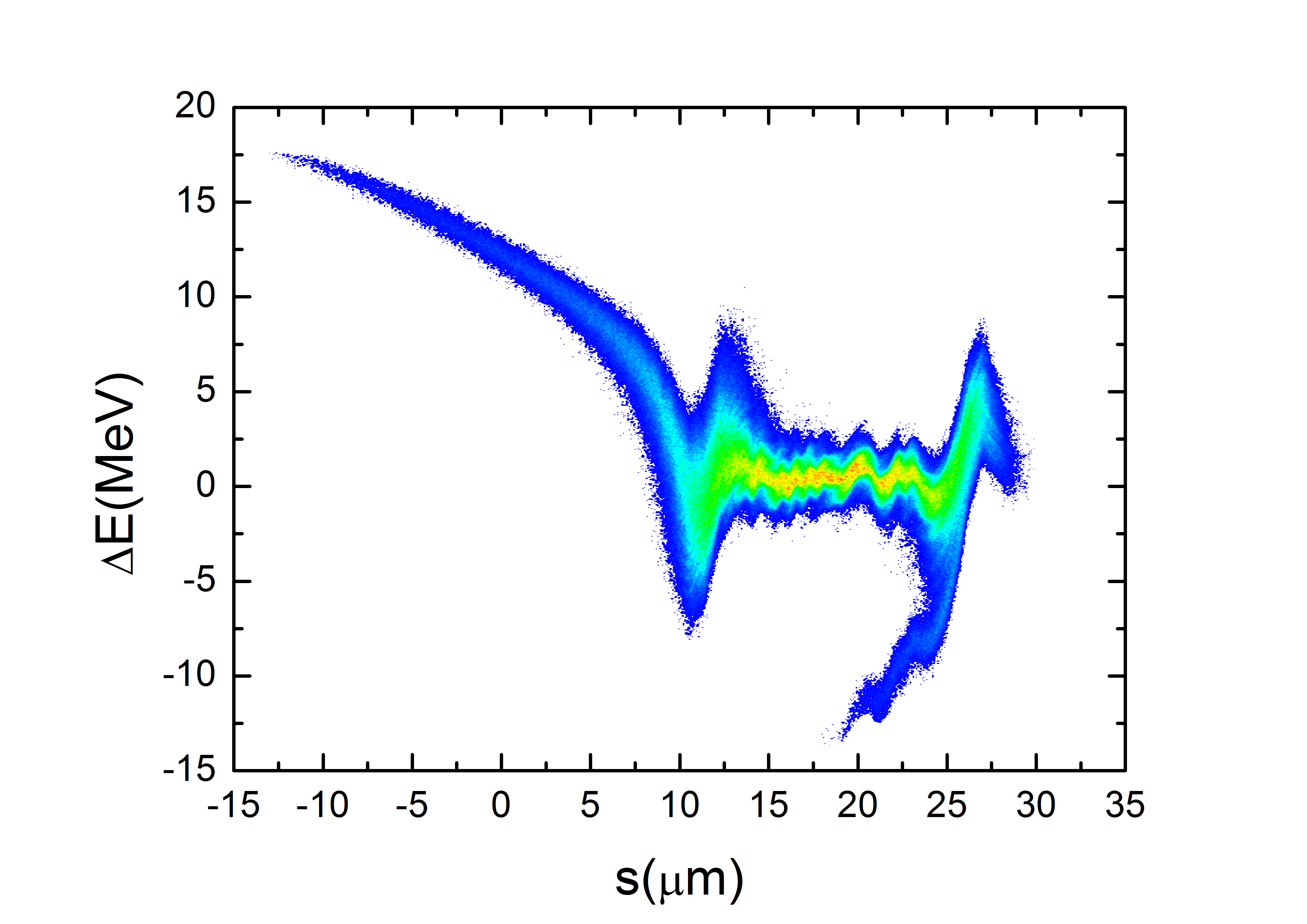}\\
  \caption{(color online). The beam phase space at the exit of the SCLF linac.}\label{figure3}
\end{figure}

The studies in this paper are based on the parameters of FEL-\textrm{II}, which is aimed to cover the photon energy of 0.4-3.0 keV by 40 segments main undulator with undulator period of 68 mm and segment length of 4 m. In order to obtain the fully coherent radiation at soft X-ray region, the baseline design of FEL-\textrm{II} is two-stage seeded FEL scheme. Fig.~\ref{figure3} shows the beam phase space at the exit of the SCLF linac. The electron beam dynamics simulation in the photon-injector is carried out by ASTRA \cite{Flottmann2003Recent} with space charge effects taken into account. ELEGANT \cite{Borland2000ELEGANT} is then used for simulation in the reminder of linac. However, from the beam phase space currently available, it is difficult to establish the two-stage seeding mode around 1 keV photon energy. In order to ensure the fully coherence, a better, i.e., more flat and uniform beam phase space should be exploited by reducing the bunch compression factor, and/or an alternative seeding mode should be prepared for backup. Therefore the idea here is that, the upstream 18 segments undulator works in SASE mode, and the downstream 22 segments undulator are seeded by the atomic XRL.

\begin{figure}
  \centering
  \includegraphics[width=8cm]{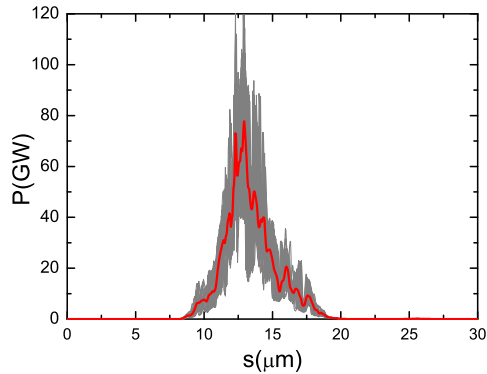}\\
  \caption{(color online). Temporal power profile at the end of undulators (91.26 m) for the SASE XFEL radiation in the beam head. Gray curves refer to 20 runs and red curves are their averages.}\label{figure4}
\end{figure}

 The XFEL simulations are normally carried out by GENESIS \cite{Reiche1999GENESIS}. 20 start-to-end time-dependent runs have been conducted and the generated power profile at the end of undulators (91.26 m) are shown Fig.~\ref{figure4}. The gray lines displays 20 separate runs, while the red line represents the average value. The SASE XFEL is able to generate nearly 77 GW soft X-ray with 8.6 fs (FWHM) pulse duration. As expected for SCLF, we simulated $4.8\times10^{12}$ photons at 1 keV photon energy.

This XFEL shoot into a gas target, an elongated plasma column is created by inner-shell photoionization within the first few femtoseconds of the XFEL pump \cite{rohringer2007x}. We establish a population inversion of the $K_\alpha$ transition in singly ionized neon at 1.46 nanometres (corresponding to a photon energy of 849 eV) created by irradiation of a gas medium. The core excited ions Auger decay within a few femtoseconds and a transient population inversion of femtosecond duration is established. This inversion forms the basis for an X-ray lasing transition. The gain of the atomic inner-shell XRL depends on the linewidth of the lasing transition. Since the inner-shell photoionization happens on an ultrafast time scale of a few femtoseconds, the ion temperature in the atomic column is expected to remain close to room temperature during the time of amplification. This opens the pathway to very narrow high-gain lasing transitions, with their width being determined only by lifetime broadening of the lower and upper lasing states. The geometry of the atomic inner-shell XRL is determined by the XFEL and at a spot size, a gain region several millimeters in length can be achieved. Due to the small aspect ratio, lasing in a single transverse mode is expected. The beam divergence has the same magnitude as the divergence of XFEL beam. Model calculations demonstrate that saturation of the XRL can be achieved at moderate gas densities.

Amplification of spontaneous emitted X-rays on the XRL transition in the exponential gain region is determined by the small-signal gain. The small-signal gain per atom is defined as \cite{rohringer2009atomic,rohringer2009atomic,darvasi2014optical,elton2012x}

\begin{align}
g(t)=\sigma_{stim}N_U(t)-\sigma_{abs}N_L(t)
\end{align}
Here $N_U(t)$ and $N_L(t)$ are occupancies of the upper and lower lasing states, and $\sigma_{stim}$ and $\sigma_{abs}$ denote the cross section for stimulated emission and absorption
\begin{align}
\sigma_{stim}=A_{UL}\frac{2\pi c^2}{\omega_{XRL}^2\Delta\omega_{XRL}},\sigma_{abs}=\frac{g_U}{g_L}\sigma_{stim}
\end{align}
where $A_{UL}$ is the Einstein $A$ coefficient for the radiative transition and $g_U$ and $g_L$ are the statistical weights of the upper and lower lasing levels. Equation (2) gives the cross sections at the peak of the line, supposing a Lorentzian line shape. The linewidth of the transition $\Delta\omega$ is dominated by the total lifetime of the upper and lower states (Auger lifetime and radiative lifetime). In the case of the Ne$^{1+}$ $2p^{-1}-1s^{-1}$ transition ($\omega$=849 eV), this results in a relative width of $\Delta\omega/\omega$=2.9$\times$10$^{-4}$ \cite{rohringer2009atomic}, where we assumed shell averaged values for Auger and radiative decay rates. The occupancies $N_U$(t) and $N_L$(t) are determined by influence of the pumping radiation on an ensemble of single atoms. The occupancies of different configuration states are calculated by solving a system of rate equations, describing valence and core photoionization, Auger, and radiative decay.


We now discuss the expected output properties of an XFEL pumped inner-shell XRL. To simulate the output of the lasing transition with highest gain, we apply a one-dimensional model that couples the atomic level kinetics to the laser propagation and amplification.

\begin{figure}
  \centering
  \includegraphics[width=8cm]{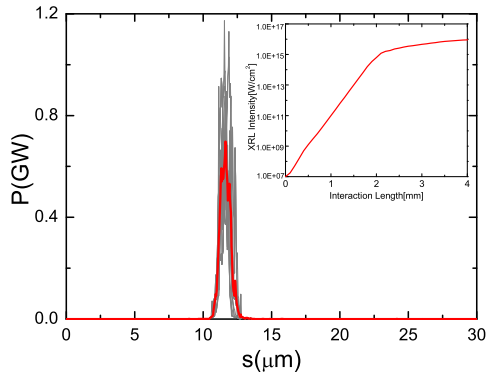}\\
  \caption{(color online). Temporal power profile of the atomic inner-shell XRL for the Ne$^{1+}$ 1s$^{-1}-2p^{-1}$ line at the exit of the neon gas. Gray curves refer to 20 runs and red curves are their averages. The inset shows output instensity of the XRL for the Ne$^{1+}$ $1s^{-1}-2p^{-1}$ line as function of the interaction length.}\label{figure5}
\end{figure}

The XRL intensity \emph{I(z,t)} is influenced by the occupancies of the upper and lower levels, which determine the small-signal gain, in the course of the propagation in the medium. Specically, light with the XRL transition frequency is attenuated for a negative small-signal gain and it is amplified for a positive small-signal gain as described by \cite{rohringer2009atomic,rohringer2009atomic,darvasi2014optical}
\begin{eqnarray}
\frac{dI_{XRL}(z,t)}{dt}&=&I_{XRL}(z,t)cn_A[\sigma_{stim}N_U(z,t)-\sigma_{abs}N_L(z,t)]\nonumber\\
                         &&+\frac{\Omega(z)}{4\pi}A_{2p\rightarrow 1s}N_U(z,t)n_Ac-c\frac{dI_{XRL}}{dz}
\end{eqnarray}
\begin{figure*}[t]
  \centering
  \subfigure{\includegraphics[width=5cm]{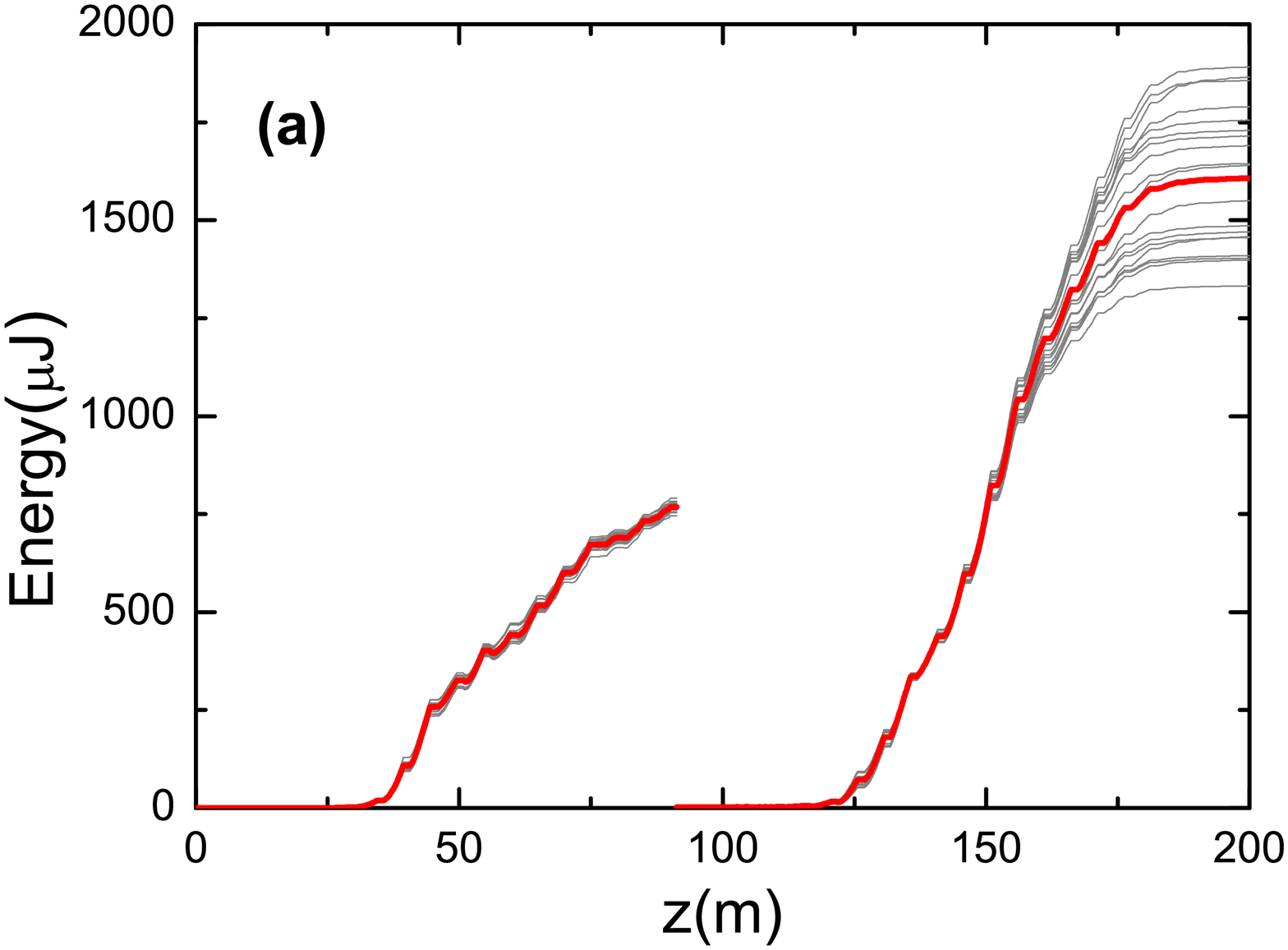}}
  \subfigure{\includegraphics[width=5cm]{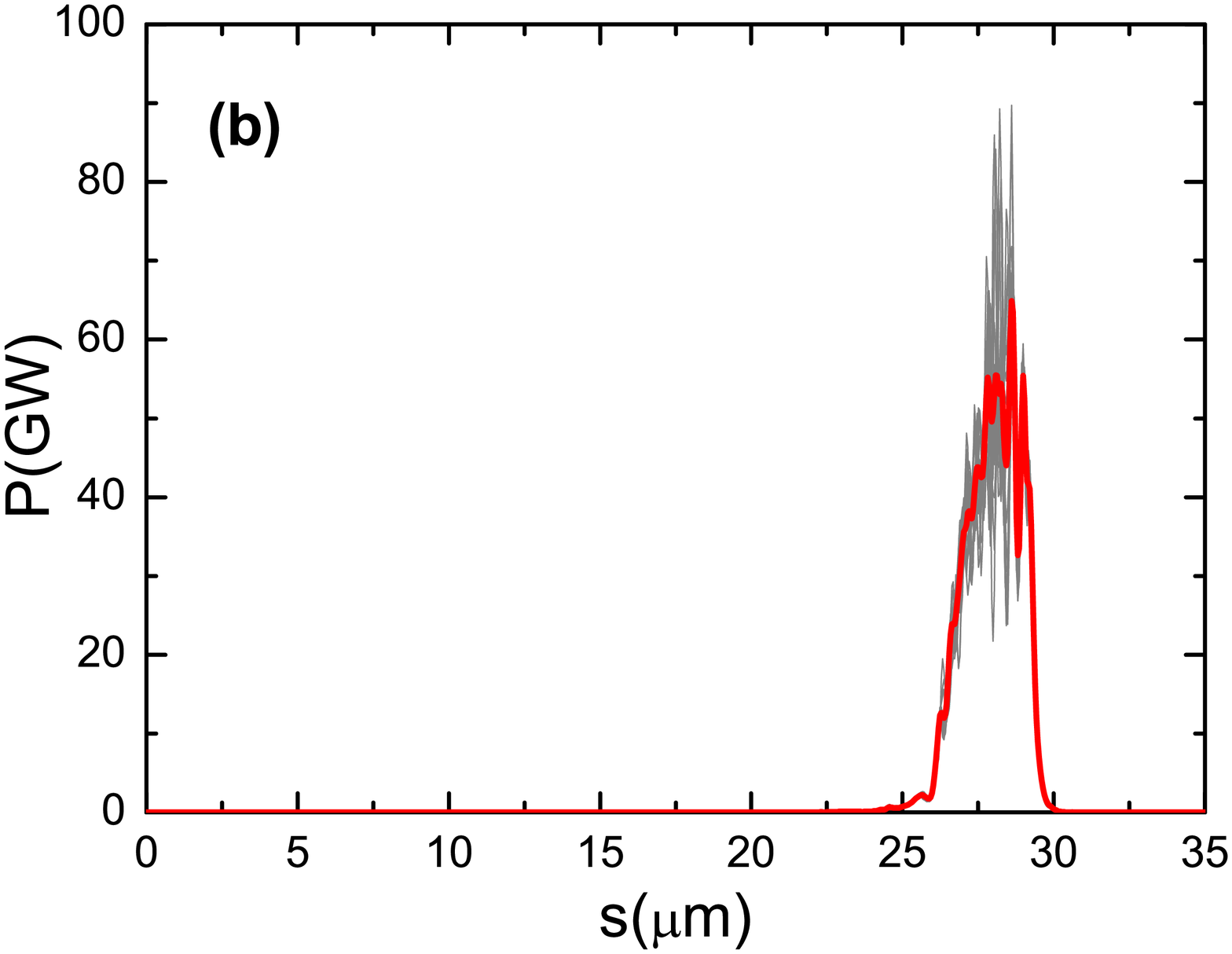}}
  \subfigure{\includegraphics[width=5cm]{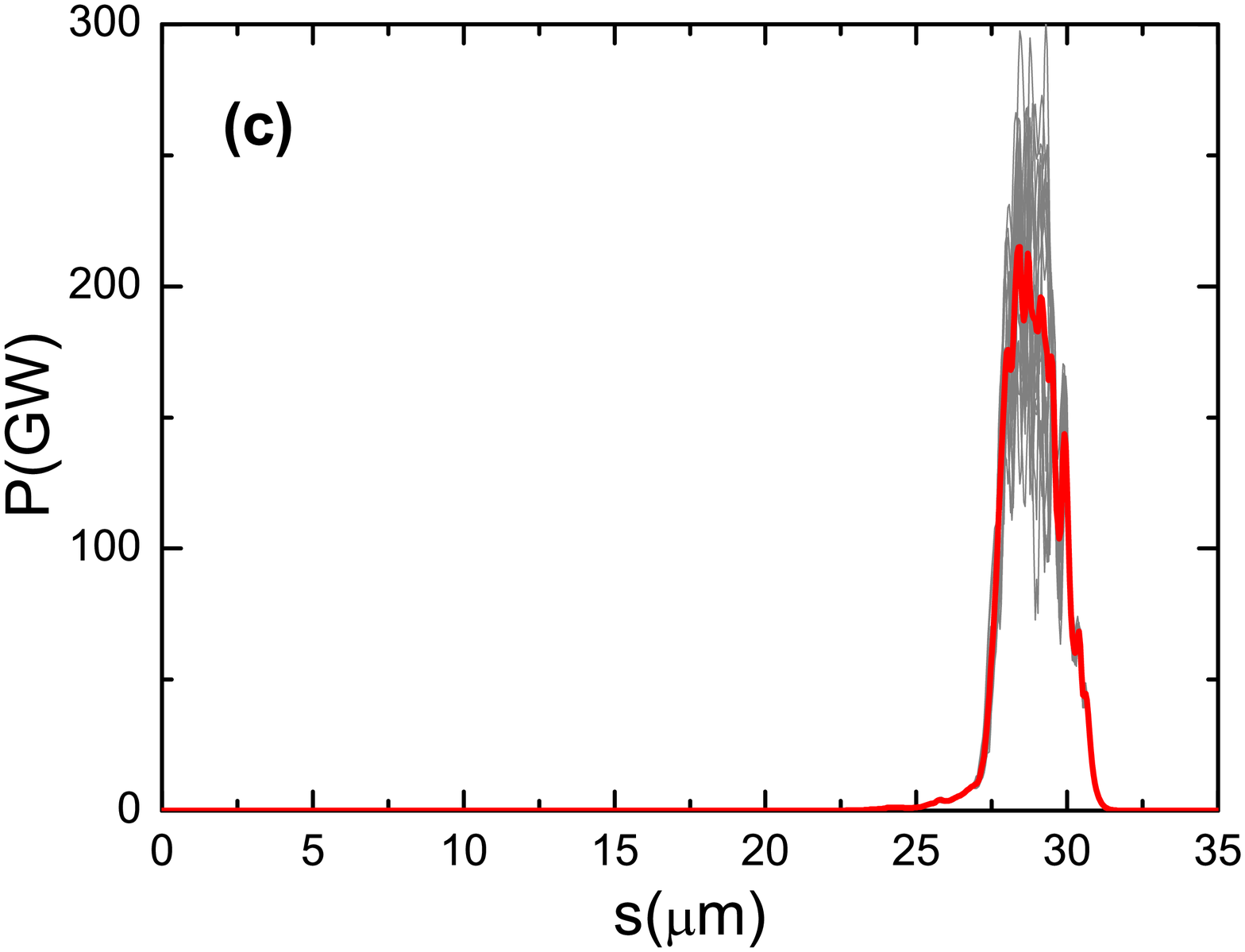}}

  \subfigure{\includegraphics[width=5cm]{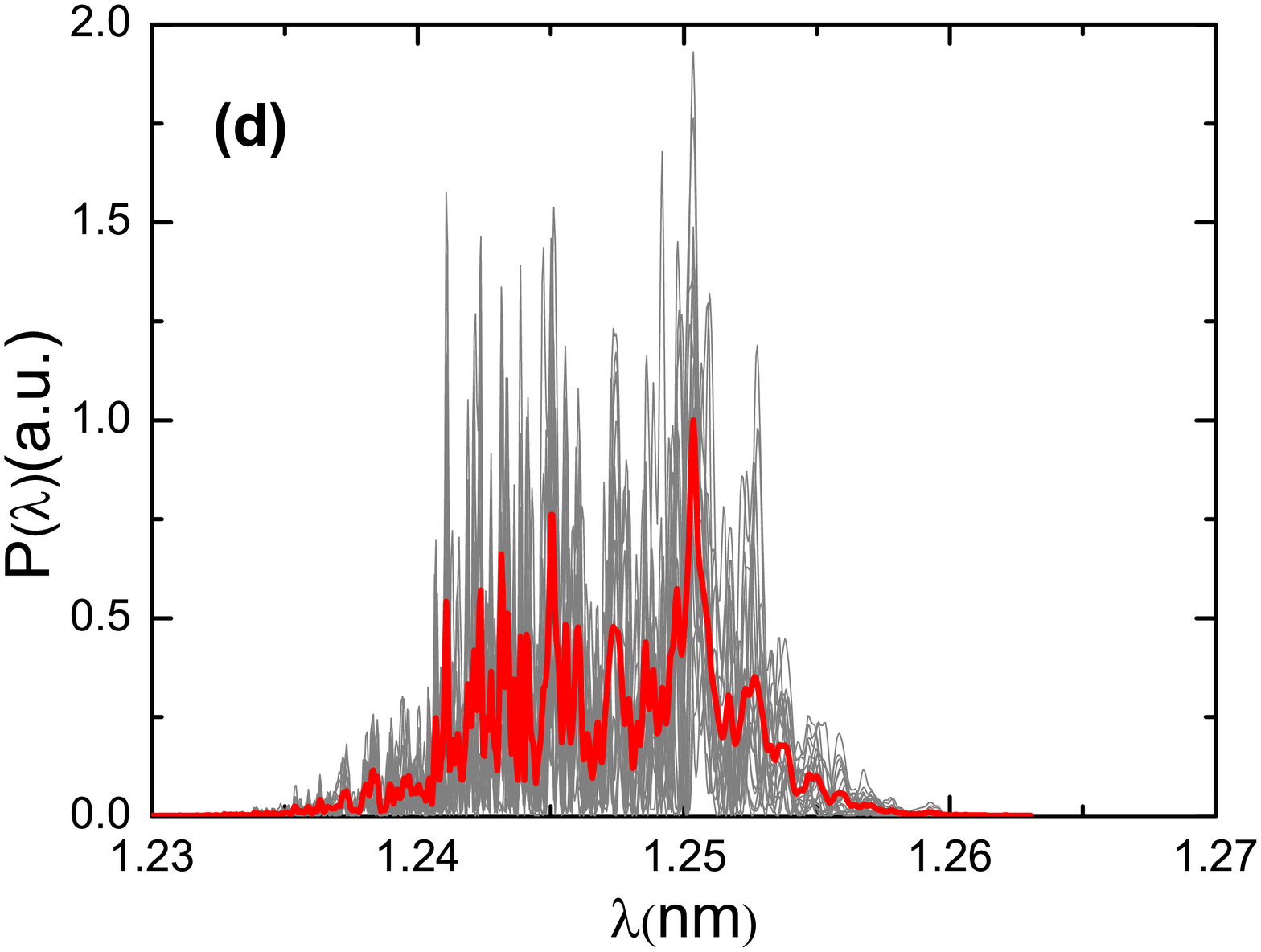}}
  \subfigure{\includegraphics[width=5cm]{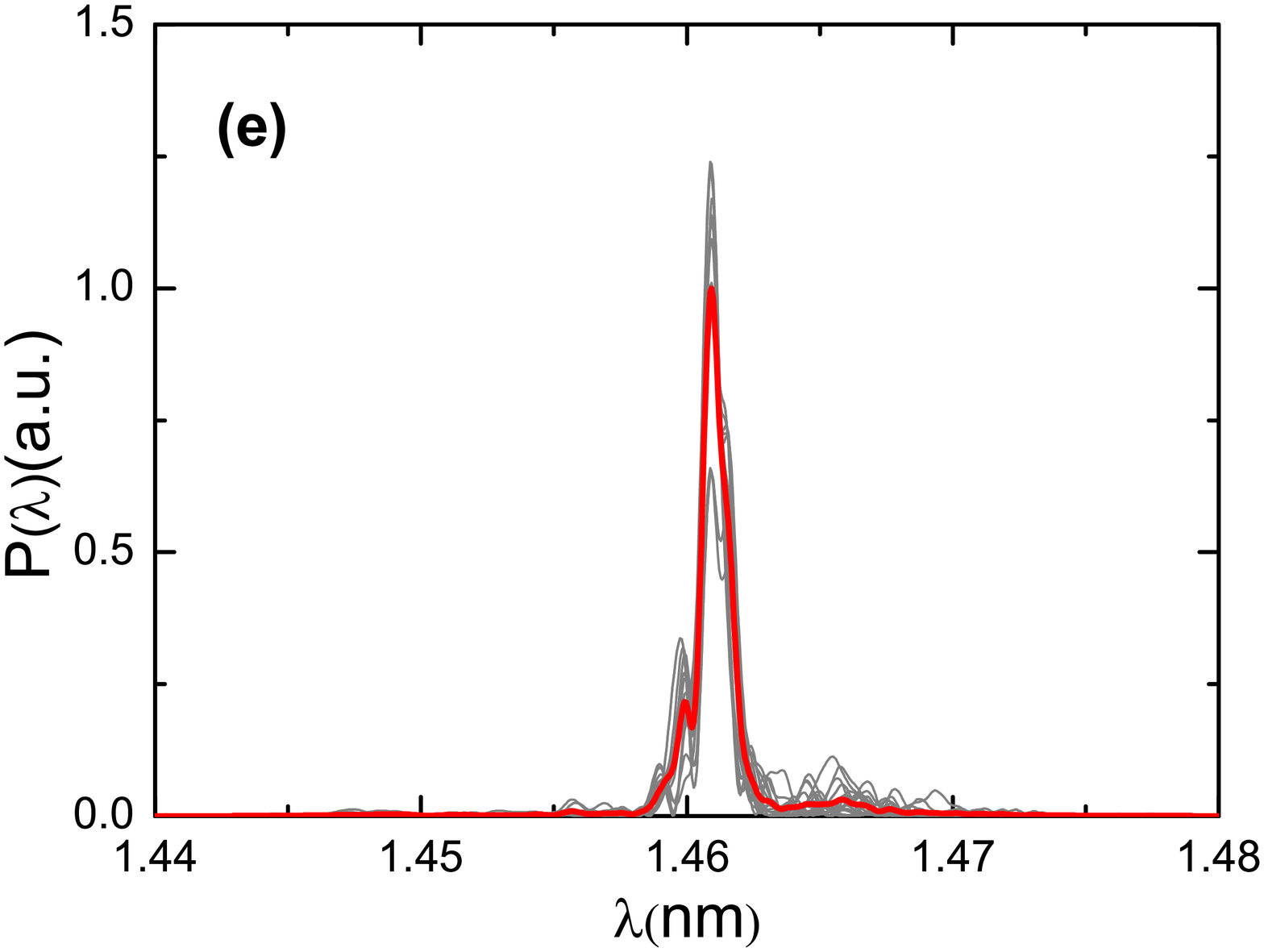}}
  \subfigure{\includegraphics[width=5cm]{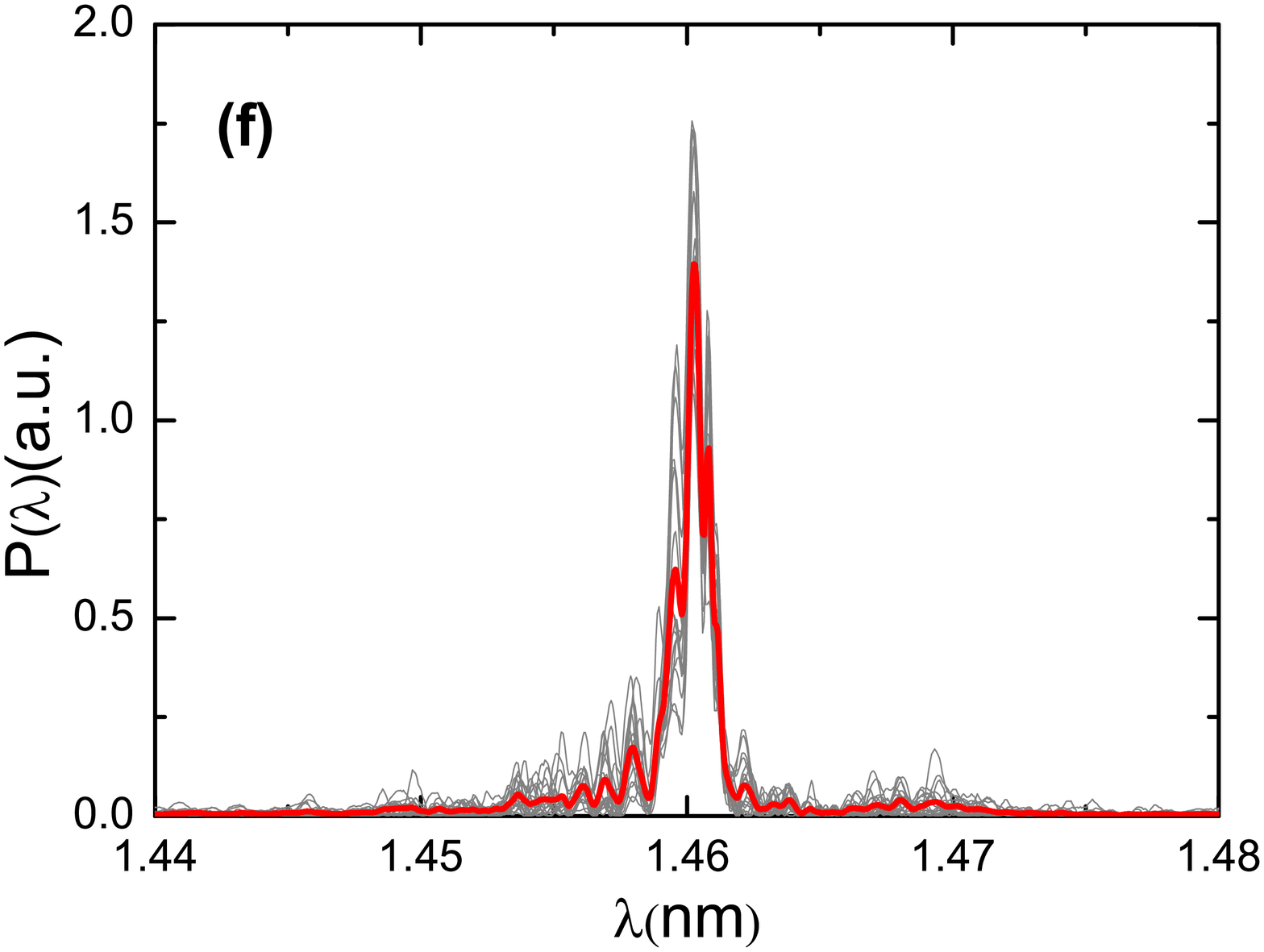}}
 \caption{\label{figure6} (color online).  20 start-to-end time-dependent runs of the calculation results, the gray lines displays 20 separate runs, while the red line represents the average value. (a) Radiation energy along z in the SASE XFEL and seeded XFEL stage . (b) Temporal power profile at the middle of undulators ($z$=50 m). (c) Temporal power profile at the end of undulators ($z$=111.54 m). The peak power of seeded XFEL can reach about 210 GW. (d) The spectrum at $z$=91.26 m for the SASE XFEL radiation in the beam head. (e) The spectrum at $z$=50 m for the seeded XFEL. (f) The spectrum at $z$=111.54 m for the seeded XFEL. The bandwidth of the SASE radiation is 7.58 eV and the bandwidth of the seeded XFEL radiation is 0.51 eV in $z$=50 m and 0.58 eV in 111.54 m, respectively.}
\end{figure*}
where $\Omega(z)=2\pi[1-(L/2-z)/\sqrt{r^2+(L/2-z)^2}]$ are the geometrical acceptance angles, allowing propagation in forward directions, $n_A$ is the atomic density, and $L$ is the interaction length and $r$ is the diameter spot. Fourth-order Runge-Kutta algorithm \cite{Milonni2010Laser} is used to solve these first-order ordinary differential equations. Due to the longitudinal pumping, lasing occurs only in forward direction. Results for the power profile at the exit are presented in Fig.~\ref{figure5}. The inner-shell XRL is able to generate nearly 0.7 GW soft X-ray with 2.6 fs (FWHM) pulse duration. Conversion efficiency is 0.91\%. The inset showing the peak intensity of the Ne$^{1+}$ lasing transition as a function of the interaction length for a neon gas density of 10$^{18}$ cm$^{-3}$.

\section{SEEDED FEL PERFORMANCE}

In the previous section, we obtained an atomic inner-shell laser that uses SASE XFEL generated by electron beam tail to pump neutral neon gas atoms. Our self-consistent gain calculations indicate that average peak power of 0.7 GW radiation can be achieved with femtosecond pulse duration. The electron beam is delayed by 18 $\upmu$m for making the heading electron overlap with seed laser. To extract more energy from electron bunch, tapered undulator technique is used \cite{kroll1981free}. The magnetic field of 22 segments downstream undulator are finely optimized by Genetic Algorithm  \cite{deb2002fast} in a steady-state GENESIS simulation. In the optimization, the magnetic field in each undulator segment are set to be equal, i.e., a stepped taper is used.

In order to reveal the real operation of the scheme, 20 start-to-end time-dependent runs have been conducted. The radiation energy along the undulator length $z$ from SASE XFEL to seeded XFEL is shown in Fig.~\ref{figure6}(a). The red curve is the average of the gray curves. The energy of the red curve at the end is 1.6 mJ. We can observe an unstable growing of the seeded XFEL power for different cases, this rms energy jitter is about 10.8\%. The output soft X-ray pulse-to-pulse power fluctuation is mainly due to the mismatch between the XFEL power and the setting of the taper.

The generated power profile at the middle and exit of the downstream undulator are shown in Fig.~\ref{figure6}(b) and Fig.~\ref{figure6}(c), respectively. In Fig.~\ref{figure6}, we also compared three spectrum distributions between the SASE radiation, and the seeded FEL radiation at the middle and at the end of the undulator line. The gray lines displays 20 separate runs, while the red line represents the average value. The seeded FEL at the end of undulators ($z$=111.54 m) is able to generate nearly 210 GW soft X-ray with 7.3 fs (FWHM) pulse duration. The bandwidth is 0.58 eV (FWHM) which corresponds to time-bandwidth product 1.02. The seeded FEL at the middle of undulators ($z$=50 m) is able to generate nearly 65 GW soft X-ray with 5.7 fs (FWHM) pulse duration. The bandwidth is 0.51 eV (FWHM) which corresponds to time-bandwidth product 0.7, at this point, seeded XFEL is almost fully coherent. In the following undulators, the SASE backgrounds from electron beam without seeding broaden the bandwidth and the microbunching structures in the electron beam contribute to the spectrum pedestal \cite{Zhang2016Microbunching,Zhang2018Eliminating}. All in all, taking advantage of the overwhelming 0.7 GW fully coherent seed laser, longitudinal coherence is enhanced at the end of undulators by a factor of 15 with respect to SASE XFEL. Note that the temporal duration of seed laser is nearly 2.6 fs, which is shorter than electron beam.

\section{conclusion}
In this paper, a scheme that combines self-seeding method and atomic inner-shell XRL to improve the longitudinal coherence of XFEL is proposed. First, a high peak power SASE is generated by the electron beam tail in the upstream undulators. Then this radiation is used to pump a given neutral neon atom, and thus produce an extremely longitudinal coherent inner-shell XRL. Finally, the electron beam is delayed properly to overlap with seed laser with its heading electrons. It is worth stressing that, while the fresh-slice technique is used for principle illustration in this paper, the proposal can be easily established with in the typical whole bunch self-seeding and/or two bunch self-seeding schemes \cite{ding2010two}.

Tapered downstream undulators are employed to enhance the output power of seeded soft X-ray. The start-to-end simulations were conducted with parameters from SCLF, which indicates that our scheme is feasible to generate 210 GW nearly fully coherent soft X-ray pulse with 7.3 fs (FWHM) pulse duration. The bandwidth is 0.58 eV (FWHM) which corresponds to time-bandwidth product 1. Its rms energy jitter is about 10.8\% at the end of the undulator. The seeding scheme is expected to deliver fully coherent radiation to beamline users and enable various experiments. Note that the scheme can be generalized to heavier atoms and therefore even shorter wavelengths in the future. In addition, the new scheme does not use crystals and gratings, which may avoid the thermal loading effects of self-seeding schemes in high repetition rate FEL facilities. Thus, the photon energy absorption and thermal lasing issue in this scheme will be seriously considered in future. An XRL generation and measurement experiment can be accomplished at the experimental station of Shanghai soft X-ray FEL user facility \cite{zhao2011shanghai}, with a focused X-ray beam size of 2 $\upmu$m$ \times $3 $\upmu$m.

\begin{acknowledgments}
The authors would like to thank M. Zhang for providing LINAC beam dynamic results of SCLF. This work was partially supported by the National Natural Science Foundation of China (11775293), the National Key Research and Development Program of China (2016YFA0401900), the Young Elite Scientist Sponsorship Program by CAST (2015QNRC001) and Ten Thousand Talent Program.
\end{acknowledgments}


\bibliographystyle{apsrev4-1}
\bibliography{paper}

\end{document}